\pdfoutput=1

\documentclass[11pt,twoside,a4paper,cmspaper,final,collab]{cms-tdr}
\def\svnVersion{493450P}\def\cmsCernNoTag{CERN-EP-2018-297}\def\cmsCernDate{\today}\def\cmsMessage{Published in the European Physical Journal C as \href{http://dx.doi.org/10.1140/epjc/s10052-019-6800-x}{\doi{10.1140/epjc/s10052-019-6800-x.}}}

\begin{document}\cmsNoteHeader{SUS-17-008}

\hyphenation{had-ron-i-za-tion}
\hyphenation{cal-or-i-me-ter}
\hyphenation{de-vices}
\RCS$HeadURL: svn+ssh://svn.cern.ch/reps/tdr2/papers/SUS-17-008/trunk/SUS-17-008.tex $
\RCS$Id: SUS-17-008.tex 490213 2019-02-26 13:03:59Z millet $

\newlength\cmsFigWidth
\newlength\cmsTabSkip\setlength{\cmsTabSkip}{2ex}
\ifthenelse{\boolean{cms@external}}{\setlength\cmsFigWidth{0.98\columnwidth}}{\setlength\cmsFigWidth{0.49\textwidth}}
\ifthenelse{\boolean{cms@external}}{\providecommand{\cmsLeft}{upper\xspace}}{\providecommand{\cmsLeft}{left\xspace}}
\ifthenelse{\boolean{cms@external}}{\providecommand{\cmsRight}{lower\xspace}}{\providecommand{\cmsRight}{right\xspace}}

\newcommand{\PSGmL}{\ensuremath{\PSGm_{\mathrm{L}}}\xspace}
\newcommand{\sNuMu}{\ensuremath{\widetilde{\nu}_{\mu}}\xspace}
\newcommand{\lpmuo}{\ensuremath{{\lambda^{\prime}_{211}}}\xspace}
\newcommand{\mzero}{\ensuremath{m_{0}}\xspace}
\newcommand{\monehalf}{\ensuremath{m_{1/2}}\xspace}
\newcommand{\mSmu}{\ensuremath{m_{\PSGm}}\xspace}
\newcommand{\mSnuMu}{\ensuremath{m_{\sNuMu}}\xspace}
\newcommand{\mNeutr}{\ensuremath{m_{\PSGczDo}}\xspace}
\newcommand{\MSlep}{\ensuremath{m(\mu_{1}\mu_{2}+\text{jets})}\xspace}
\newcommand{\MChi}{\ensuremath{m(\mu_{2}\text{j}_{1}\text{j}_{2})}\xspace}
\newcommand{\wpwp}{\ensuremath{\Wpm\Wpm}\xspace}
\newcommand{\VV}{\ensuremath{\mathrm{V}\mathrm{V}}\xspace}
\newcommand{\WZ}{\ensuremath{\PW\PZ}\xspace}
\newcommand{\ZZ}{\ensuremath{\PZ\PZ}\xspace}
\newcommand{\VH}{\ensuremath{\mathrm{V}\PH}\xspace}
\newcommand{\VVV}{\ensuremath{\mathrm{V}\mathrm{V}\mathrm{V}}\xspace}
\newcommand{\ttbarVH}{\ensuremath{\ttbar(\mathrm{V},\PH)}\xspace}
\newcommand{\ttbarV}{\ensuremath{\ttbar\mathrm{V}}\xspace}
\newcommand{\ttbarZ}{\ensuremath{\ttbar\PZ}\xspace}
\newcommand{\ttbarW}{\ensuremath{\ttbar\PW}\xspace}
\newcommand{\ttbarH}{\ensuremath{\ttbar\PH}\xspace}
\newcommand{\ttbargamma}{\ensuremath{\ttbar\gamma}\xspace}
\newcommand{\ttbarttbar}{\ensuremath{\ttbar\ttbar}\xspace}
\newcommand{\gammaX}{\ensuremath{\PGg+\mathrm{X}\xspace}}
\newcommand{\Vgamma}{\ensuremath{\mathrm{V}\PGg\xspace}}
\newcommand{\WWgamma}{\ensuremath{\PW\PW\PGg\xspace}}
\newcommand{\WZgamma}{\ensuremath{\PW\PZ\PGg\xspace}}
\newcommand{\tgamma}{\ensuremath{\cPqt\PGg}\xspace}
\newcommand{\tZq}{\ensuremath{\cPqt\PZ\cPq}\xspace}
\newcommand{\ggH}{\ensuremath{\Pg\Pg\PH}\xspace}
\newcommand{\dThreeD}{\ensuremath{d_{\mathrm{3D}}}\xspace}
\newcommand{\sigmadThreeD}{\ensuremath{\sigma(\dThreeD)}\xspace}
\newcommand{\miniIso}{\ensuremath{I_{\text{mini}}}\xspace}
\newcommand{\ptRatio}{\ensuremath{\pt^{\text{ratio}}}}
\newcommand{\ptRel}{\ensuremath{\pt^{\text{rel}}}\xspace}
\newcommand{\ptCorr}{\ensuremath{\pt^{\text{corr}}}\xspace}
\newcommand{\etl}{\ensuremath{\epsilon_{\mathrm{TL}}}\xspace}
\newcommand{\ptISR}{\ensuremath{\pt^{\mathrm{ISR}}}}
\providecommand{\NA}{\ensuremath{\text{---}}}
\ifthenelse{\boolean{cms@external}}{\providecommand{\cmsTable}[1]{#1}}{\providecommand{\cmsTable}[1]{\resizebox{\textwidth}{!}{#1}}}

\cmsNoteHeader{SUS-17-008}

\title{Search for resonant production of second-generation sleptons with same-sign dimuon events in proton-proton collisions at $\sqrt{s} = 13\TeV$}
\titlerunning{Search for second-generation sleptons}

\date{\today}

\abstract{
A search is presented for resonant production of second-generation sleptons (\PSGmL, \sNuMu) via the $R$-parity-violating coupling \lpmuo to quarks, in events with two same-sign muons and at least two jets in the final state. The smuon (muon sneutrino) is expected to decay into a muon and a neutralino (chargino), which will then decay into a second muon and at least two jets. The analysis is based on the 2016 data set of proton-proton collisions at $\sqrt{s}=13\TeV$ recorded with the CMS detector at the LHC, corresponding to an integrated luminosity of 35.9\fbinv. No significant deviation is observed with respect to standard model expectations. Upper limits on cross sections, ranging from 0.24 to 730\unit{fb}, are derived in the context of two simplified models representing the dominant signal contributions leading to a same-sign muon pair. The cross section limits are translated into coupling limits for a modified constrained minimal supersymmetric model with \lpmuo as the only nonzero $R$-parity violating coupling. The results significantly extend restrictions of the parameter space compared with previous searches for similar models.
}

\hypersetup{
pdfauthor={CMS Collaboration},
pdftitle={Search for resonant production of second-generation sleptons with same-sign dimuon events in proton-proton collisions at sqrt(s) = 13 TeV},
pdfsubject={CMS},
pdfkeywords={CMS, SUSY, supersymmetry, RPV, R-parity-Violation}}
\maketitle

\section{Introduction}
\label{sec:introduction}
Supersymmetry (SUSY)~\cite{Ramond:1971gb,Golfand:1971iw,Neveu:1971rx,Volkov:1972jx,Wess:1973kz,Wess:1974tw,Fayet:1974pd,FARRAR1978575,BURAS197866,PhysRevD.26.287,Nilles:1983ge,Haber:1984rc,Martin:1997ns} is an attractive extension of the standard model (SM) offering gauge coupling unification and a solution to the hierarchy problem.
In SUSY, a symmetry between fermions and bosons is postulated that assigns a new fermion (boson) to every SM boson (fermion). These new particles are called superpartners or sparticles.
The superpotential of a minimal SUSY theory can contain lepton and baryon number violating terms~\cite{PhysRevD.26.287},
\begin{linenomath}
\ifthenelse{\boolean{cms@external}}
{
\begin{multline}
W_{\text{RPV}} = \frac12 \lambda_{\mathrm{ijk}} L_{\mathrm{i}} L_{\mathrm{j}} \overline{E}_{\mathrm{k}}  + \lambda^{\prime}_{\mathrm{ijk}} L_{\mathrm{i}} Q_{\mathrm{j}} \overline{D}_{\mathrm{k}} - \kappa_{\mathrm{i}} L_\mathrm{i} H_\mathrm{u}\\
+ \frac{1}{2} \lambda^{\prime\prime}_{\mathrm{ijk}} \overline{U}_{\mathrm{i}} \overline{D}_{\mathrm{j}} \overline{D}_{\mathrm{k}}.
  \label{eq:superpotential}
\end{multline}
}
{
\begin{equation}
W_{\text{RPV}} = \frac12 \lambda_{\mathrm{ijk}} L_{\mathrm{i}} L_{\mathrm{j}} \overline{E}_{\mathrm{k}}  + \lambda^{\prime}_{\mathrm{ijk}} L_{\mathrm{i}} Q_{\mathrm{j}} \overline{D}_{\mathrm{k}} - \kappa_{\mathrm{i}} L_\mathrm{i} H_\mathrm{u} + \frac{1}{2} \lambda^{\prime\prime}_{\mathrm{ijk}} \overline{U}_{\mathrm{i}} \overline{D}_{\mathrm{j}} \overline{D}_{\mathrm{k}}.
  \label{eq:superpotential}
\end{equation}
}
\end{linenomath}
Here, $\mathrm{i},\mathrm{j},\mathrm{k}\in\{1,2,3\}$ are generation indices, $L$, $Q$ and $H_{\mathrm{u}}$ are the lepton, quark, and up-type Higgs $SU(2)_{\mathrm{L}}$ doublet superfields, respectively, and $\overline{E}$, $\overline{D}$, $\overline{U}$ are the charged lepton, down-type quark, and up-type quark $SU(2)_{\mathrm{L}}$ singlet superfields, respectively. The $SU(2)_{\mathrm{L}}$ weak isospin and $SU(3)_{\mathrm{C}}$ color indices are suppressed. The terms associated with the coupling parameters $\lambda$, $\lambda^{\prime}$, and $\kappa$ would lead to lepton number violation, while the one linked to $\lambda^{\prime\prime}$ would cause baryon number violation.
A combination of these terms would lead to a rapid decay of the proton, which is not observed. To preserve the proton stability, additional symmetries are introduced.
A common choice is to introduce $R$-parity conservation (RPC), which forbids all the terms in Eq.~(\ref{eq:superpotential}).
The $R$-parity of a particle is defined as $(-1)^{2s+3(B-L)}$~\cite{FARRAR1978575}, where $s$, $B$, and $L$ denote the spin, the baryon number, and the lepton number of the particle, respectively.
However, there are other symmetries that can replace $R$-parity and keep the proton stable~\cite{baryon-triality-1,baryon-triality-2}.
SUSY theories in which $R$-parity conservation is not imposed are usually called $R$-parity violating (RPV) models. A detailed review of RPV SUSY can be found in Ref.~\cite{Barbier:2004ez}.
In RPC SUSY models, sparticles can only be produced in pairs, and the lightest sparticle (LSP) is stable.
If the LSP is neutral (\eg, the lightest neutralino \PSGczDo), experimental signatures at hadron colliders usually involve a large amount of missing transverse momentum due to undetected LSPs.
In RPV SUSY models, the signatures can differ greatly from RPC scenarios. The LSP can decay back into SM particles, and the strong exclusion limits for sparticles from RPC searches do not necessarily apply to RPV models.
In addition, RPV models allow for different production mechanisms, such as the resonant production of sleptons from $\cPq\cPaq$ collisions, which will be investigated in this paper.

{\tolerance=800
At the CERN LHC, sleptons---the scalar superpartners of leptons---can be produced in $\cPq\cPaq$ interactions as $s$-channel resonances via the trilinear $LQ\overline{D}$ term of the superpotential. The coupling strength of this interaction is characterized by $\lambda^{\prime}_{\mathrm{ijk}}$, where $\mathrm{i}$ specifies the lepton and $\mathrm{j}, \mathrm{k}$ the quark generations. For proton-proton ($\Pp\Pp$) collisions at the LHC, the contributions from the first quark generation ($\mathrm{j}=\mathrm{k}=1$) are dominant. The lepton index determines which sleptons can be produced via this coupling. It also defines the possible decay modes of the LSP, since all decay modes of the LSP into SM particles must involve RPV couplings.
Resonant slepton production was first proposed in Refs.~\cite{DIMOPOULOS1988210,PhysRevD.41.2099,DREINER1991597} as a viable signature for RPV SUSY at hadron colliders. Detailed studies of resonant slepton production leading to a same-sign (SS) dilepton signature were presented in Refs.~\cite{PhysRevD.63.055008,Deliot2001,PhysRevD.75.035003}. Resonant slepton production was also suggested as a possible explanation for observed deviations from the SM at the Tevatron and the LHC~\cite{PhysRevD.84.055012,PhysRevD.91.015011,PhysRevD.93.115022}.
\par}

{\tolerance=800
This paper focuses on the resonant production of second-generation sleptons (\PSGmL, \sNuMu) via the RPV coupling \lpmuo in final states with an SS muon pair and jets.
The search is based on $\sqrt{s} = 13\TeV$ $\Pp\Pp$ collision data recorded in 2016 with the CMS detector at the LHC, corresponding to an integrated luminosity of 35.9\fbinv.
Limits on resonant production of second-generation sleptons were set by the \DZERO collaboration \cite{d0_lp211} at the Fermilab Tevatron and in Ref.~\cite{PhysRevD.86.055010} reinterpreting ATLAS and CMS results.
The results presented in this paper are the first bounds on resonant slepton production in this channel set by CMS.
Assuming RPC, searches for pair production of charged sleptons exclude slepton masses up to 450\GeV for {\PSe} and \PSGm~\cite{Sirunyan:2018nwe} and 500\GeV for \PSe, \PSGm, and \PSGt~\cite{Aaboud:2018jiw} if the left- and right-handed sleptons are mass degenerate and assuming a massless LSP. For the production of left-handed smuons only, the exclusion limits decrease to 280\GeV~\cite{Sirunyan:2018nwe}. Searches for SUSY scenarios with two SS leptons and jets in the final state have been performed by ATLAS~\cite{Aaboud:2017dmy} and CMS~\cite{Sirunyan:2017uyt} using $\Pp\Pp$ collision data recorded in 2016 without finding any evidence for physics beyond the SM. While the search presented in Ref.~\cite{Sirunyan:2017uyt} targets various RPC SUSY signals, this paper focuses on RPV SS dimuon signatures from resonant slepton production. The main experimental differences are related to the definition of the signal regions (SRs), the momentum thresholds for the muons, and the fact that no lower bound on the missing transverse momentum is applied here. A recent review of searches and bounds on RPV SUSY can be found in Ref.~\cite{Dercks:2017lfq}.
\par}

\begin{figure}[!htbp]
\centering
\includegraphics[width=0.45\textwidth]{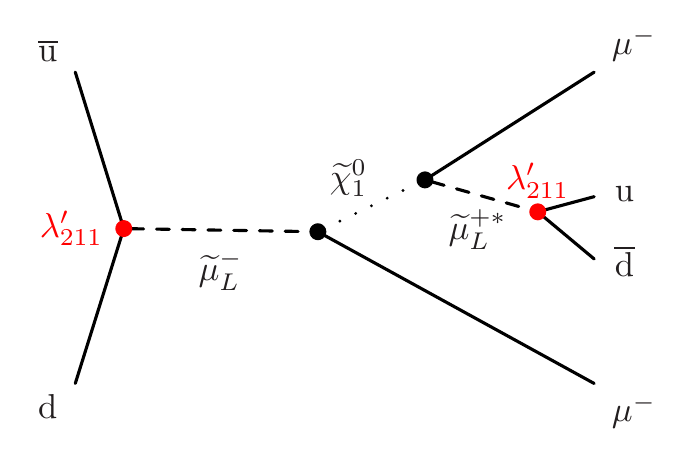}
\includegraphics[width=0.45\textwidth]{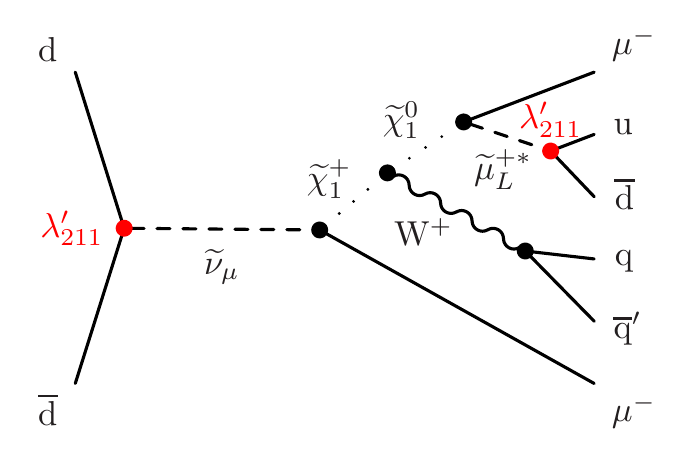}
  \caption{Signal contributions from a modified cMSSM with \lpmuo as an additional coupling, which are considered as simplified signal models SM1 (\cmsLeft) and SM2 (\cmsRight) in this search. The charge conjugate diagrams are included as well.}
\label{fig:smfeyn}
\end{figure}

Based on a modified version of the constrained minimal SUSY model (cMSSM)~\cite{PhysRevD.49.6173} with \lpmuo as an additional coupling, two of the dominant signal processes leading to an SS muon pair are shown in Fig.~\ref{fig:smfeyn}. Here, the LSP is assumed to be the lightest neutralino \PSGczDo, and all other RPV couplings are set to zero (single-coupling dominance). In the diagrams shown in Fig.~\ref{fig:smfeyn}, a smuon (\PSGmL) or a muon sneutrino (\sNuMu) is produced in $\cPq\cPaq$ ($\cPqu\cPaqd$, $\cPaqu\cPqd$, $\cPqd\cPaqd$) annihilation and decays into a muon and either the LSP neutralino (\PSGczDo) or the lightest chargino (\PSGcpmDo). The \PSGcpmDo will further decay into the LSP and a {\PW} boson. All decay chains in Fig.~\ref{fig:smfeyn} end with the decay of the LSP into a second muon and two light quarks via an off-shell smuon ($\PSGmL^{*}$) in an effective three-body decay. The decay of the $\PSGmL^{*}$ involves the RPV coupling \lpmuo, so that $R$-parity is violated in the production and the decay of the slepton.
The probed values of \lpmuo are large enough to ensure a prompt decay of the LSP.
Because of the Majorana nature of the LSP, the second muon will have the same charge as the first one with a probability of 50\%. Same-sign dilepton production is rare in the SM, and is therefore well suited as a signature for new physics searches.

For the signal models, a simplified model approach \cite{Alves:2011wf,Chatrchyan:2013sza} is used, where the dominant signal contributions are extracted and simulated as independent signals assuming a branching fraction of 100\%. One advantage of this approach is that the final exclusion limits are less model dependent than for one based strictly on the cMSSM, since the sparticle masses can be set to combinations not allowed in the cMSSM, and the signal contributions are split into the different production mechanisms and decay chains. The \cmsLeft and \cmsRight diagrams of Fig.~\ref{fig:smfeyn} will be called simplified model 1 (SM1) and simplified model 2 (SM2), respectively. Another important contribution to SS muon pair production via \lpmuo in the modified cMSSM comes from a process similar to the one shown in Fig.~\ref{fig:smfeyn} (\cmsRight). In this process, a \PSGmL is produced and decays as $\PSGmL \to \PSGczDt \mu$ (instead of $\sNuMu \to \PSGcpmDo \mu$). The \PSGczDt then decays into a {\PZ} boson and the LSP. As long as the {\PW} boson from Fig.~\ref{fig:smfeyn} (\cmsRight) and the {\PZ} boson decay into quarks, there is no difference in analysis sensitivity between these processes. Therefore, exclusion limits of SM2 will also apply for this additional decay chain. The results of the search are interpreted in terms of SM1 and SM2 as well as the modified cMSSM.

\section{The CMS detector and event reconstruction}
The central feature of the CMS apparatus is a superconducting solenoid of 6\unit{m} internal diameter, providing a magnetic field of 3.8\unit{T}. Within the solenoid volume are a silicon pixel and strip tracker, a lead tungstate crystal electromagnetic calorimeter (ECAL), and a brass and scintillator hadron calorimeter (HCAL), each composed of a barrel and two endcap sections. Forward calorimeters extend the pseudorapidity ($\eta$) coverage provided by the barrel and endcap detectors. Muons are detected in gas-ionization chambers embedded in the steel flux-return yoke outside the solenoid. A more detailed description of the CMS detector, together with a definition of the coordinate system used and the relevant kinematic variables, can be found in Ref.~\cite{Chatrchyan:2008zzk}. Events of interest are selected using a two-tiered trigger system~\cite{Khachatryan:2016bia}. The first level, composed of custom hardware processors, uses information from the calorimeters and muon detectors to select events at a rate of around 100\unit{kHz} within a time interval of less than 4\mus. The second level, known as the high-level trigger, consists of a farm of processors running a version of the full event reconstruction software optimized for fast processing, and reduces the event rate to around 1\unit{kHz} before data storage.

The particle-flow algorithm~\cite{CMS-PRF-14-001} aims to reconstruct and identify each individual particle in an event, with an optimized combination of information from the various elements of the CMS detector. The energy of electrons is determined from a combination of the electron momentum at the primary interaction vertex as determined by the tracker, the energy of the corresponding ECAL cluster, and the energy sum of all bremsstrahlung photons spatially compatible with originating from the electron track. The energy of charged hadrons is determined from a combination of their momentum measured in the tracker and the matching ECAL and HCAL energy deposits, corrected for zero-suppression effects and for the response function of the calorimeters to hadronic showers. Finally, the energy of neutral hadrons is obtained from the corresponding corrected ECAL and HCAL energies. The missing transverse momentum vector \ptvecmiss is defined as the projection onto the plane perpendicular to the beam axis of the negative vector sum of the momenta of all reconstructed particle-flow objects in an event. Its magnitude is referred to as \ptmiss.

Hadronic jets are clustered from these reconstructed particles using the infrared and collinear safe anti-\kt algorithm~\cite{Cacciari:2008gp, Cacciari:2011ma} with a distance parameter of 0.4. The jet momentum is determined as the vectorial sum of all particle momenta in the jet, and is found from simulation to be within 5 to 10\% of the true momentum over the whole transverse momentum (\pt) spectrum and detector acceptance~\cite{Khachatryan:2016kdb}. Additional proton-proton interactions within the same or nearby bunch crossings can contribute additional tracks and calorimetric energy depositions to the jet momentum. To mitigate this effect, tracks identified to be originating from pileup vertices are discarded, and an offset factor is applied to correct for remaining contributions. Jet energy corrections are derived from simulation to bring the measured response of jets to that of particle level jets on average. In situ measurements of the momentum balance in dijet, photon+jet, \PZ+jet, and multijet events are used to account for any residual differences in jet energy scale in data and simulation. Additional selection criteria are applied to each jet to remove jets potentially dominated by anomalous contributions from various subdetector components or reconstruction failures. Jets are classified as originating from a bottom quark ({\cPqb} tagged) if they pass the medium working point requirements from the combined secondary vertex algorithm (v2)~\cite{Sirunyan:2017ezt}. The medium working point is defined to have a misidentification probability of 1\% for jets from light quarks or gluons in a simulated multijet sample. For this working point, the {\cPqb} jet identification efficiency is around 63\% for {\cPqb} jets with $\pt>20\gev$ in simulated \ttbar events.

{\tolerance=800
Muons are measured in the range $\abs{\eta} < 2.4$, with detection planes made using three technologies: drift tubes, cathode strip chambers, and resistive plate chambers. Matching muons to tracks measured in the silicon tracker results in a relative \pt resolution, for muons with \pt up to 100\GeV, of 1\% in the barrel and 3\% in the endcaps. The \pt resolution in the barrel is better than 7\% for muons with \pt up to 1\TeV~\cite{Sirunyan:2018}.
\par}

{\tolerance=800
The reconstructed vertex with the largest value of summed physics-object $\pt^2$ is taken to be the primary $\Pp\Pp$ interaction vertex. The physics objects are the jets, clustered using the anti-\kt jet finding algorithm~\cite{Cacciari:2008gp,Cacciari:2011ma} with the tracks assigned to the vertex as inputs, and the associated missing transverse momentum, taken as the negative vector sum of the \pt of those jets. More details are given in Section~9.4.1 of Ref.~\cite{CMS-TDR-15-02}.
\par}

\section{Monte Carlo simulation}
\label{sec:simulation}
{\tolerance=800
The \MGvATNLO~\cite{Alwall:2014hca} v2.2.2 generator is used to simulate the following background processes: \wpwp, \ttbarV, \Vgamma, \WWgamma, \WZgamma, \tgamma, \ttbargamma, \VVV, \VH, \ttbarttbar, and \tZq ($\mathrm{V}=\PW,\PZ$). Except for the \wpwp process that is simulated at leading order (LO)~\cite{Alwall:2007fs,Alwall:2008qv,Kalogeropoulos:2018cke} accuracy, the simulations are done at next-to-leading order (NLO)~\cite{Frederix:2012ps} accuracy in terms of perturbative quantum chromodynamics (QCD) and include up to one or two additional partons at the matrix element level. The simulations for \WZ, \ZZ, \ttbarH, and \ggH are generated with \POWHEG v2~\cite{Nason:2004rx,Frixione:2007vw,Alioli:2010xd,Melia:2011tj,Nason:2013ydw,Hartanto:2015uka,Bagnaschi:2011tu} at NLO accuracy. Simulations of double parton scattering leading to the production of \PW\PW{} are done with \PYTHIA~v8.205~\cite{Sjostrand:2007gs}. The parton showering and hadronization is simulated using \PYTHIA~v8.212 with the \textsc{CUETP8M1}~\cite{Skands:2014pea,CMS-PAS-GEN-14-001} tune for the underlying event. Double counting of additional partons between \MGvATNLO and \PYTHIA is removed with the appropriate technique for each simulation (MLM matching for LO~\cite{Alwall:2007fs,Alwall:2008qv}, FxFx merging for NLO~\cite{Frederix:2012ps}). All samples include a simulation of the contributions from pileup that is matched to the data with a reweighting technique. The parton distribution functions (PDFs) are NNPDF3.0 LO~\cite{nnpdf30} for LO and NNPDF3.0 NLO~\cite{nnpdf30} for NLO samples, respectively. The \GEANTfour~\cite{Geant} package is used to model the detector response for all background processes.
\par}

Monte Carlo (MC) simulated signal samples are produced for both simplified models defined in Section \ref{sec:introduction} using \MGvATNLO at LO accuracy with NNPDF3.0 LO PDFs and \PYTHIA for hadronization and showering. The detector simulation makes use of the CMS fast simulation package~\cite{Abdullin:2011zz}.
The mass scans range from 200 to 3000\GeV for the slepton mass, and from 100 to 2900\GeV for the LSP mass, with a 100\GeV spacing. For SM2, the mass of the chargino is calculated from the LSP and slepton mass as follows, using three different values of $x$ (0.1, 0.5, 0.9):
\begin{linenomath}
\begin{equation}
\label{eq:sm2x}
m_{\PSGcpmDo} = \mNeutr + x \left( \mSnuMu - \mNeutr \right).
\end{equation}
\end{linenomath}
For SM2, some points of the scans are omitted since the mass difference between the LSP and \PSGcpmDo would force the \PW\ boson to be off-shell.
All signal studies and simulations are based on the MSSM-RpV-TriRpV model implementation in the \textsc{sarah}~\cite{Staub:2013tta,Staub:2011dp,Goodsell:2014bna,Goodsell:2016udb,Staub:2017jnp} package. For the full model interpretation within the modified cMSSM, mass spectra and branching fractions have been calculated with the \textsc{SPheno}~\cite{Porod:2003um,Porod:2011nf} package.

\section{Event selection}
\label{sec:sel}
Events with the targeted signal signature will have exactly two muons with the same electric charge, at least two jets from light quarks (\cPqu{}, \cPqd), and no jets from \cPqb{} quarks.
The following event requirements are designed to efficiently select signal-like events while rejecting SM background.
Events are selected using triggers that require at least one muon candidate with $\pt > 50\GeV$ within $\abs{\eta} < 2.4$.
Typical trigger efficiencies for muons passing the identification criteria described below are around 90\%.

Events are selected with exactly two well-identified muons within the acceptance of $\abs{\eta} < 2.4$. The \pt of the leading (subleading) muon is required to be larger than 60 (20)\GeV. In addition, the two muons are required to have the same electric charge and to have a dimuon invariant mass larger than 15\GeV. The muon reconstruction relies on the results of a global fit using measurements from the silicon tracker as well as the muon detectors. For muon candidates to be well identified, the global fit is required to be consistent with the measurements of the individual subsystems, and the relative uncertainty in the measured muon \pt is required to be smaller than 0.2.

{\tolerance=800
To ensure that muon candidates originate from the primary vertex, the impact parameter, and the longitudinal displacement from the primary vertex of the corresponding point on the trajectory must be smaller than 0.5 and 1\mm, respectively. The ratio $\abs{\dThreeD}/\sigmadThreeD$ is required to be smaller than 4, where \dThreeD is the three-dimensional impact parameter with respect to the primary vertex and \sigmadThreeD its uncertainty from the track fit.
\par}

The isolation criterion for muons is based on a combination of three variables (\miniIso, \ptRatio, \ptRel) and is designed to provide an efficient selection of muons from heavy-particle decays (\eg, \PW\ and \PZ\ bosons, and sparticles) especially in systems with a high Lorentz boost, where decay products and jets may overlap \cite{Khachatryan:2016kod}.

{\tolerance=1500
The mini isolation (\miniIso) is defined as the scalar sum of the \pt of neutral hadrons, charged hadrons, and photons inside a cone of $\Delta R = \sqrt{(\Delta\eta)^2 + (\Delta\phi)^2}$ (where $\phi$ is the azimuthal angle in radians) around the muon direction at the vertex, divided by the muon \pt. The cone size depends on the lepton \pt as
\par}
\begin{linenomath}
\begin{equation}
\Delta R \left(\pt(\ell)\right) = \dfrac{10\GeV}{\min\left[\max\left(\pt(\ell), 50\GeV\right), 200\GeV\right]}.
\end{equation}
\end{linenomath}
The varying isolation cone helps to reduce the inefficiency from accidental overlap between the muon and jets in a busy event environment.
The second isolation variable (\ptRatio) is defined as the ratio of the muon \pt and the \pt of the closest jet within $\Delta R = 0.4$ around the muon. The \ptRel variable is then defined as the transverse momentum of the muon with respect to that jet after subtracting the muon:
\begin{linenomath}
\begin{equation}
\ptRel=\frac{\abs{\left[\vec{p}(\text{jet})-\vec{p}\left(\ell\right)\right] \times \vec{p}(\ell)}}{\abs{\vec{p}(\text{jet})-\vec{p}(\ell)}}.
\end{equation}
\end{linenomath}
If no jet is found within $\Delta R < 0.4$, \ptRatio{} (\ptRel) is set to 1 (0). Muons are classified as isolated if they fulfill the requirements
\begin{linenomath}
\begin{equation}
\miniIso < 0.16 \text{\ \ and\ \ } (\ptRatio > 0.76 \text{\ \ or\ \ } \ptRel > 7.2\GeV).
\end{equation}
\end{linenomath}
Events are required to have at least two jets with $\pt > 40\GeV$ and $\abs{\eta} < 2.4$. Jets that do not pass a set of quality criteria or are within $\Delta R < 0.4$ of a lepton are not included in this count. The quality criteria are designed to reject jets that are likely to originate from anomalous energy deposits~\cite{Khachatryan:2014gga}. Events with one or more \cPqb-tagged jets fulfilling the criteria listed above, but with a lowered \pt threshold of 30\GeV, are rejected. This requirement helps in reducing background from \ttbar events as well as contributions from \ttbarV and \ttbarH production.

Several additional event vetoes are applied to reduce contributions from multilepton backgrounds. Events with additional muons, one or more electrons, or hadronically decaying tau leptons are rejected. For the muon veto a looser set of identification criteria is used. In addition, the \pt threshold is lowered to 5\GeV, and the isolation criterion is replaced by $\miniIso < 0.4$. Electron identification is based on track quality, the shape of the energy deposits in the ECAL, and the ratio of energy deposits in the HCAL and ECAL. Electron candidates with missing hits in the innermost tracking layers or those assigned to a photon conversion are rejected. As an additional criterion, the mini isolation variable for electron candidates (similarly defined as for muons) is required to be smaller than 0.4. All electrons with $\pt > 10\GeV$, $\abs{\eta} < 2.5$, and fulfilling the criteria described above are used for the electron veto. Hadronically decaying \Pgt{} candidates are reconstructed with the hadron-plus-strips algorithm and identified with a decay mode finding algorithm selecting one- and three-prong decays \cite{Khachatryan:2015dfa}. The candidates that fulfill the identification criteria, $\pt > 30\GeV$, and $\abs{\eta} < 2.3$, are used for the \index{tau}tau lepton veto.

To further separate signal and background, the SR is divided into ten bins indicated by SR1 to SR10 in the plane of \MSlep and \MChi, where \MSlep is defined as the invariant mass of the two muons and all selected jets in the event, and \MChi is the invariant mass of the subleading muon and the two leading jets. Events from signal processes would lead to a broad peak around the slepton mass along the \MSlep axis. The expected shape of the signal in \MChi depends on the involved masses. While SM1 yields a broad peak around the LSP mass in the \MChi distribution for the vast majority of mass combinations, the peak for SM2 signals tends to be shifted to higher masses if one of the particles entering the \MChi calculation is not from the LSP decay. The SR binning is chosen such that each signal will typically only contribute to a very small number of SR bins. The bins range from 0--500, 500--1000, 1000--1500 and $>$1500\GeV in both variables and are numbered in ascending order starting from the bins with an \MChi of 0--500\GeV and increasing with \MSlep.

\section{Background estimation}
\label{sec:backgrounds}

The sources of the SM background contributions can be divided into three classes: processes with two prompt muons, with at least one nonprompt muon, and with at least one muon from an internal photon conversion.

{\tolerance=800
Processes with two prompt SS muons are estimated with MC simulation. The dominant contributions with prompt leptons come from \WZ and SS \wpwp production. The contributions from \WZ, \wpwp, and \ZZ are labeled as VV in the following. Other important backgrounds arise from \ttbar in association with a \PW, \PZ, or Higgs boson (\ttbarVH). All additional contributions with two prompt SS muons are labeled as "other" and include \VVV, \ttbarttbar, \tZq, \VH, \ggH, and double parton scattering processes. The normalization of the \WZ and \ttbarZ processes is derived from a fit to data using the distribution of the number of \cPqb-tagged jets in a control region (CR) with three muons, at least two jets, and $\ptmiss > 30\GeV$. Two of the three muons are required to have opposite sign and invariant mass within 15\GeV around the \PZ\ boson mass. This results in scale factors to the simulation-based \WZ and \ttbarZ estimates of $1.22 \pm 0.15$ and $1.15 \pm 0.50$, respectively. All additional prompt background estimates are based on simulation only. For \WZ events with three prompt muons from the \PW{} and \PZ{} decay, an additional correction is applied to correct for potential differences in the third lepton veto efficiency between data and simulation.
\par}

Contributions from events with at least one nonprompt muon are estimated with the tight-to-loose ratio method. These events arise mostly from \ttbar production, where one of the muons is produced in the decay of a bottom hadron. The tight-to-loose ratio method has two main steps. First, the ratio of the number of muons passing the tight working point to the number of muons passing the loose one (\etl) is measured in a CR that is dominated by SM events consisting of jets produced through the strong interaction (QCD multijet events). Here, tight muons are muons fulfilling all selection criteria from Section \ref{sec:sel}, while loose muons have relaxed constraints on the isolation. This measurement region contains events with exactly one loose muon candidate and at least two jets. To reduce the contamination of prompt leptons in the \etl measurement (mostly from $\PW \to \mu \nu$), the transverse mass of the lepton and {\ptmiss} for events in the CR has to be smaller than 30\GeV. The remaining contribution from prompt leptons is estimated from simulation and subtracted from the numerator and denominator of \etl. Typical values for \etl are in the range of 0.05--0.07. In the second step, events from application regions are used as a proxy for the nonprompt contributions to the SR. Events in these regions have to pass the same requirements as SR events, with the exception that one or both muons fulfill only the loose, but not the tight, selection criteria. The contributions from events with two prompt muons are removed using simulations. For each muon that is loose but not tight the event is weighted with $\etl/(1-\etl)$. The measurement of \etl is performed as a function of muon $\eta$ and \ptCorr, which is defined as the muon \pt corrected according to the amount of energy in the isolation cone above the tight threshold. This is done to reduce the impact of differences between the measurement region (QCD multijet dominated) and the application regions (\ttbar dominated). A detailed explanation of the tight-to-loose ratio method and the definition of \ptCorr is given in Refs.~\cite{Khachatryan:2016kod,Sirunyan:2017uyt}.

Another source of SM background is due to internal photon conversion, where a virtual photon converts into two muons. If the decay is very asymmetric, only one of the muons will pass the muon \pt threshold. Such conversions combined with the production of, \eg, a \PW\ boson can contribute to the SR. The performance of the conversion background simulation is validated in a three-lepton CR, where the invariant mass of the opposite-sign muon pair closest to the \PZ\ boson mass ($m_{\PZ}$) is smaller than 75\GeV and the invariant mass of the three muons fulfills $\abs{m_{\mu\mu\mu}-m_{\PZ}} < 15\GeV$. The resulting yields in data and simulation are consistent within the normalization uncertainty assigned to these processes (see Section \ref{sec:unc}). This background is referred to as \gammaX{} in the following.

The most important backgrounds in the first two SR bins are processes with nonprompt muons followed by \VV production.
With increasing \MChi and \MSlep, the nonprompt background contributions become less relevant, making \VV production the most important background for the other SR bins.
Nonprompt and \VV backgrounds account for 78\% of the overall background. The next most important background is \ttbarVH production making up around 10\% of the total background. The remaining 12\% originates in equal amounts from \gammaX{} and the rare processes grouped as other backgrounds.
Studies based on simulations indicate that the charge misidentification probability is negligible for muons passing the chosen identification criteria.

\begin{table*}[!ht]
\centering
\topcaption{Sources of systematic uncertainties considered in this search and the range of yield variations in the signal regions. The background uncertainties are given as fractions of the total background yields in the signal regions. For the signal, the ranges covering the most relevant signal regions for each signal are given. The first three blocks affect the background predictions and list all experimental uncertainties, uncertainties for processes where the yield is obtained from data, and additional uncertainties for simulation-based backgrounds. In the last block, additional uncertainties for the signal prediction are shown.}
\begin{tabular}{l c c}
\hline
Source                              & Background (\%) & Signal (\%)\\
\hline
Integrated luminosity                 &  1--2           &    2.5     \\
Pileup                                &  0--6           &    1--3    \\
Trigger efficiency                    &  1--2           &    1       \\
Muon selection                        &  3--6           &    6       \\
{\cPqb} tagging                       &  0--2           &    1--2    \\
Jet energy scale and resolution       &  1--8           &    1--5    \\[\cmsTabSkip]
Nonprompt muon estimate               &  0--21          &    \NA     \\
\WZ normalization                     &  1--3           &    \NA     \\
\ttbarZ normalization                 &  0--3           &    \NA     \\[\cmsTabSkip]
\wpwp normalization                   &  2--17          &    \NA     \\
\ttbarW normalization                 &  0--3           &    \NA     \\
\gammaX, other, \ttbarH normalization &  1--14          &    \NA     \\
Scale and PDF variations (shape)      &  0--9           &    0--1    \\
\wpwp generator comparison            &  0--13          &    \NA     \\
\WZ third lepton veto                 &  1--4           &    \NA     \\
Stat.~precision of simulations        &  3--32          &    \NA     \\[\cmsTabSkip]
Stat.~precision signal efficiency     &  \NA            &    1--4    \\
Initial state radiation               &  \NA            &    0--2    \\
Muon fast simulation                  &  \NA            &    4       \\
\hline
\end{tabular}
\label{tab:uncertainties}
\end{table*}
\section{Systematic uncertainties}
\label{sec:unc}

The expected yields and shapes of background and signal processes are affected by different systematic uncertainties.
The uncertainties taken into account for this search are summarized in Table~\ref{tab:uncertainties}.

Experimental uncertainties include those related to the integrated luminosity, pileup modeling, trigger efficiencies, muon identification efficiencies, {\cPqb} tagging efficiencies, and jet energy measurement. These uncertainties are taken into account for both expected signal and background yields. For the integrated luminosity measurement an uncertainty of 2.5\% is assigned~\cite{CMS-PAS-LUM-17-001}. The pileup simulation uses the total inelastic cross section, which is varied around its nominal value to obtain an uncertainty estimate. This results in shifts of 0--8\% in the expected yields for individual SR bins. The trigger, muon identification, and {\cPqb} tagging efficiencies are measured in data and in simulation. The differences between the two are corrected for by applying scale factors to the simulated events. Uncertainties in these measurements are propagated to the scale factors and used as systematic uncertainties. For the trigger efficiency measured in an independent data set this results in an uncertainty of 2\% on the predicted simulation-based background yields. The muon identification uncertainty amounts to 3\% per muon, which is based on tag-and-probe measurement techniques. For the {\cPqb} tagging efficiency~\cite{Sirunyan:2017ezt}, the scale factors vary by 1--2\% for {\cPqb} jets and around 10\% for light jets. This leads to yield variations between 1 and 2\% for simulation-based backgrounds. The jet energy measurement in simulation is corrected to match the energy scale as well as the resolution observed in data. Adding these two uncertainties in quadrature leads to variations between 1 and 8\% of the background yields from simulation. For the nonprompt muon background estimate, several uncertainties are taken into account. The statistical uncertainty due to the finite number of events in the tight-to-loose ratio measurement region and the application region is propagated to the resulting event yields. In addition, uncertainties due to prompt-lepton contamination in the tight-to-loose ratio measurement are considered. In total, this results in uncertainties between 32 and 56\% for this background. The fit to obtain the normalization of \WZ and \ttbarZ processes, described in Section \ref{sec:backgrounds}, results in scale factors with uncertainties of 15\% (50\%) for the \WZ (\ttbarZ) process, which include both statistical and systematic components.

For the main backgrounds estimated from simulation (\VV, \ttbarV), theoretical uncertainties are assessed by varying the QCD factorization and normalization scales by factors of 2 and 0.5, respectively.
The asymmetric combinations, where one of the scales is multiplied by a factor of 2 while the other is multiplied by a factor of 0.5, are omitted \cite{Cacciari:2003fi,Catani:2003zt}.
In addition, the different replicas of the NNPDF3.0~\cite{nnpdf30} set are used to estimate the uncertainties due to the proton PDFs.
This results in normalization uncertainties of 21\% (14\%) for \wpwp (\ttbarW) production. For \WZ and \ttbarZ only the difference in shape is taken into account, since the normalization and its uncertainty are obtained from data.
For the less important backgrounds (\gammaX, \ttbarH, other), a flat 50\% normalization uncertainty is used instead of the scale and PDF variations for each process group.
The uncertainties in the shapes of \VV and \ttbarV processes due to scale and PDF variations are below 10\%.
Based on a comparison of samples from different generators (\MGvATNLO, \POWHEG), an additional uncertainty is assigned to the \wpwp background estimate, which amounts to 4--25\%.
The uncertainty in the third lepton veto efficiency correction for \WZ is in the range of 7--24\% and obtained from the uncertainty in the scale factors.
For all processes, uncertainties due to limited sample sizes are taken into account. These are taken as uncorrelated among the individual SR bins and only affect the shape but not the overall expected yields. Their magnitude is within 3--32\%.

{\tolerance=800
The signal efficiencies and the corresponding uncertainties due to limited sample sizes are calculated with the Wilson score interval~\cite{wilsonscore}. Typical values of the uncertainties for SR bins with at least 5\% of the yields at a given signal point are within 1--4\%. The \MGvATNLO modeling of initial-state radiation (ISR), which affects the total transverse momentum (\ptISR) of the slepton, is improved by reweighting the \ptISR{} distribution in signal events. The reweighting procedure is based on studies of the \pt of \PZ boson events in data~\cite{isr_rew_8tev}. The reweighting factors range between 1.18 at $\ptISR = 125\GeV$ and 0.78 for $\ptISR > 600\GeV$. Their deviation from 1.0 is taken as systematic uncertainty in the reweighting.
\par}

Residual differences in the muon selection efficiencies between the CMS fast simulation package used for signal samples and the full detector simulation with \GEANTfour are corrected by applying additional scale factors. The systematic uncertainties assigned to these scale factors are 2\% per muon, resulting in a 4\% uncertainty in the signal yield.

\section{Results and interpretations}
\begin{figure}
\centering
\includegraphics[width=\cmsFigWidth]{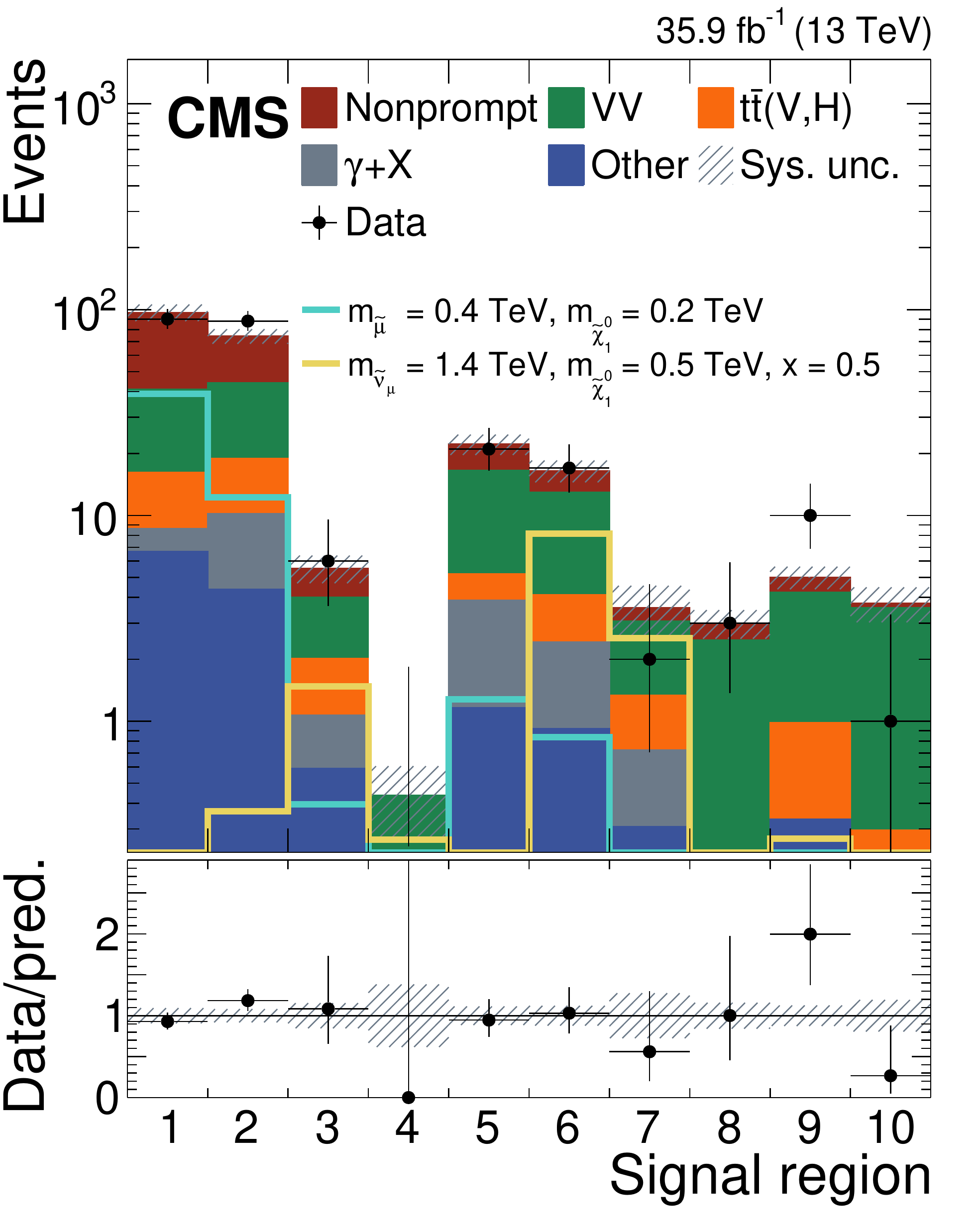}
\caption{Expected (after fit) and observed event yields in the SR bins as defined in Table~\ref{tab:eventyields}. The gray band shows the systematic uncertainty in the background yields. Also shown are the expected yields for two signal points normalized to their expected limit on the cross section. The vertical bars denote the Poisson confidence intervals calculated with the Garwood procedure, while the horizontal bars show the bin width.}
\label{fig:eventyields}
\end{figure}

\begin{figure*}[!htb]
\centering
\includegraphics[width=\cmsFigWidth]{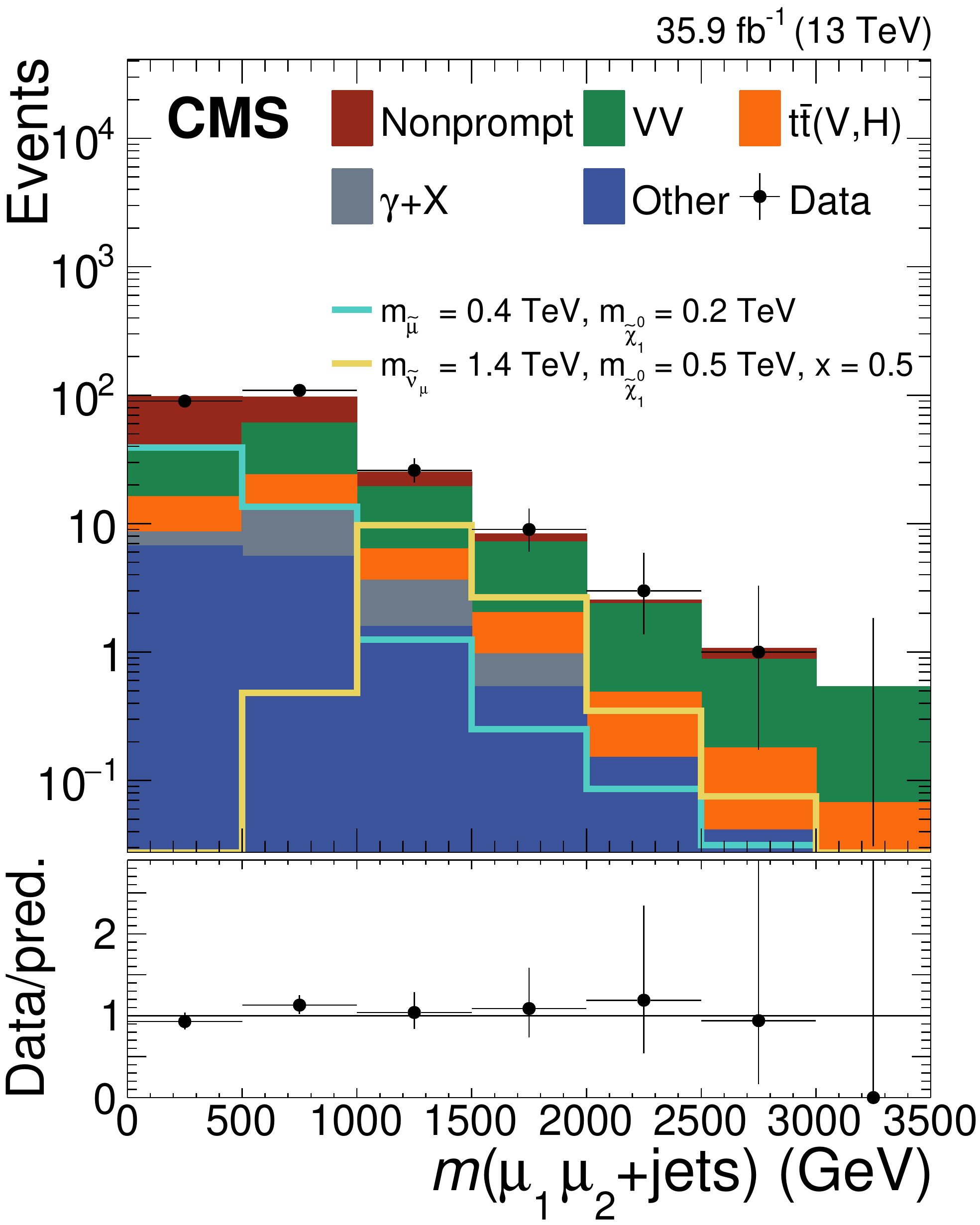}
\includegraphics[width=\cmsFigWidth]{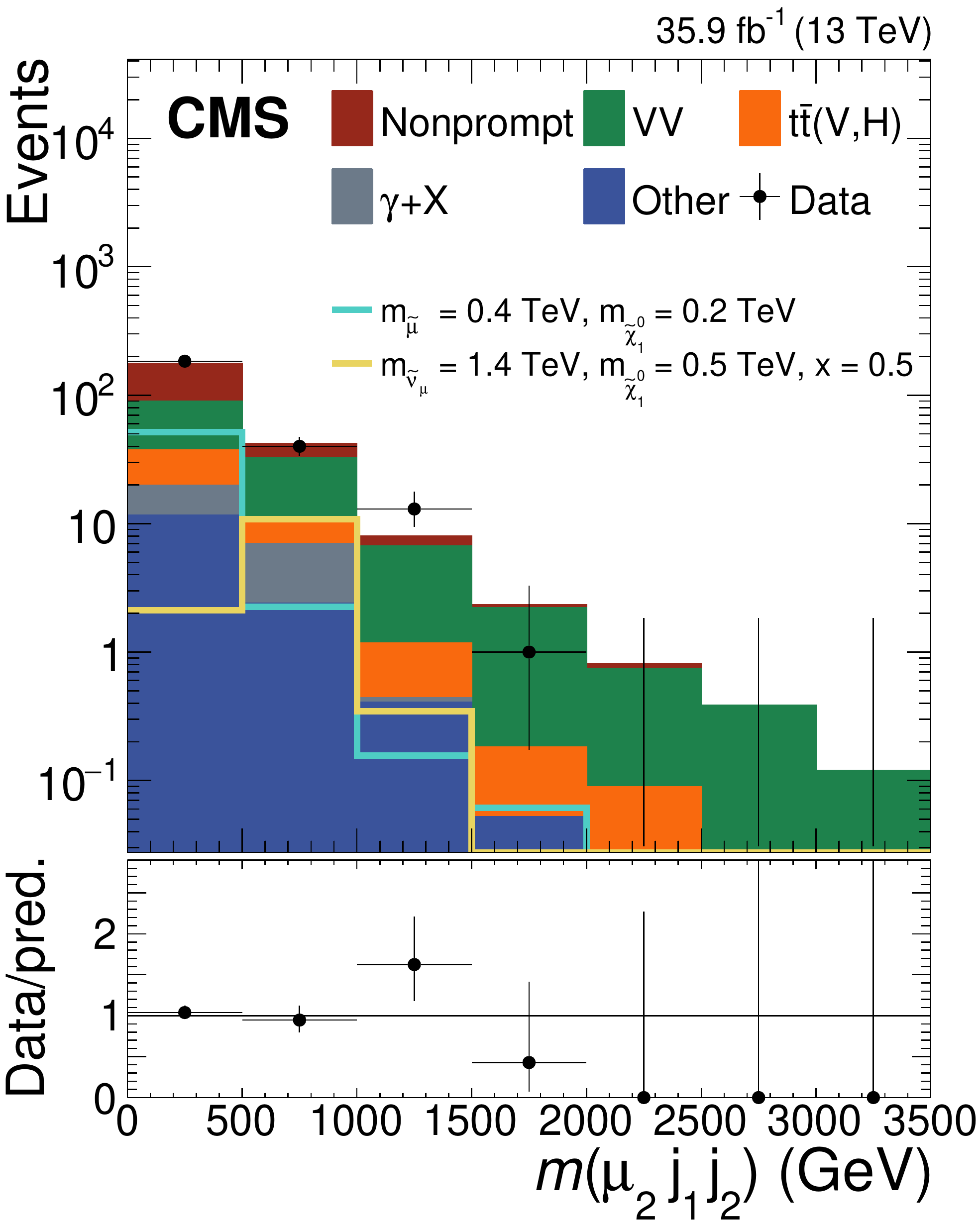}
\caption{Expected (after fit) and observed event yields in the \MSlep and \MChi distribution. Here, \MSlep is defined as the invariant mass of both muons and all jets in the event, and \MChi is the invariant mass of the subleading muon and the two leading jets. Also shown are the expected yields for two signal points normalized to their expected limit on the cross section. The vertical bars denote the Poisson confidence intervals calculated with the Garwood procedure, while the horizontal bars show the bin width.}
\label{fig:M1M2}
\end{figure*}

The expected and observed yields for the different SR bins are listed in Table \ref{tab:eventyields} and shown in Fig.~\ref{fig:eventyields}. The distributions of \MSlep and \MChi are shown in Fig.~\ref{fig:M1M2}.
For the background estimates shown in these figures, all systematic uncertainties listed in Section \ref{sec:unc} are included as nuisance parameters and constrained in a maximum likelihood fit of the expected background to the observed data assuming the background-only hypothesis. Table~\ref{tab:eventyields} shows the expected yields before and after the fit.
No significant deviation is observed with respect to SM expectations. For all signal points, the highest observed deviation from the SM expectations is 2.0 standard deviations. This deviation is observed for the SM1 signal with $\mSmu= 0.7\TeV$ and $\mNeutr = 0.3\TeV$, which has its main contribution in SR2.

\begin{table*}
\centering
\topcaption{Expected and observed event yields in the signal regions. The uncertainties are the total systematic uncertainties in the expected yields. Also shown are the expected yields for two signal points normalized to the expected limits on the cross sections.}
\cmsTable{
\begin{tabular}{cccccccc}
\hline
\multirow{3}{*}{SR} & \multirow{2.5}{*}{\MChi} & \multirow{2.5}{*}{\MSlep} & \multirow{2.5}{*}{Exp.~SM} & \multirow{2.5}{*}{Exp.~SM} & \multirow{3}{*}{Data} & SM1 & SM2 ($x=0.5$)\\
 & \multirow{2.5}{*}{({\GeVns})} & \multirow{2.5}{*}{({\GeVns})} & \multirow{2.5}{*}{(before fit)} & \multirow{2.5}{*}{(after fit)} & & $\mSmu = 0.4\TeV$& $\mSnuMu = 1.4\TeV$\\
 & & & & & & $\mNeutr = 0.2\TeV$& $\mNeutr = 0.5\TeV$\\
\hline
 1 & \multirow{4}{*}{0--500}     & 0--500     &   82.0  $\pm$   19.0  &   96.9  $\pm$    9.0  &   90 &   39.0  $\pm$    4.6  &     $<$0.01          \\
 2 &                             & 500--1000  &   62.0  $\pm$   11.0  &   74.3  $\pm$    6.0  &   88 &   12.3  $\pm$    1.7  &    0.37 $\pm$    0.06\\
 3 &                             & 1000--1500 &    4.84 $\pm$    0.99 &    5.53 $\pm$    0.85 &    6 &    0.40 $\pm$    0.11 &    1.48 $\pm$    0.19\\
 4 &                             & $>$1500    &    0.41 $\pm$    0.15 &    0.44 $\pm$    0.17 &    0 &    0.04 $\pm$    0.02 &    0.27 $\pm$    0.04\\[\cmsTabSkip]
 5 & \multirow{3}{*}{500--1000}  & 500--1000  &   19.6  $\pm$    3.5  &   22.2  $\pm$    2.5  &   21 &    1.29 $\pm$    0.22 &    0.12 $\pm$    0.02\\
 6 &                             & 1000--1500 &   14.5  $\pm$    2.6  &   16.5  $\pm$    2.0  &   17 &    0.84 $\pm$    0.16 &    8.18 $\pm$    0.94\\
 7 &                             & $>$1500    &    4.00 $\pm$    1.30 &    3.57 $\pm$    0.98 &    2 &    0.14 $\pm$    0.05 &    2.54 $\pm$    0.35\\[\cmsTabSkip]
 8 & \multirow{2}{*}{1000--1500} & 1000--1500 &    2.70 $\pm$    0.56 &    2.99 $\pm$    0.47 &    3 &    0.03 $\pm$    0.02 &    0.08 $\pm$    0.01\\
 9 &                             & $>$1500    &    4.39 $\pm$    0.78 &    5.01 $\pm$    0.63 &   10 &    0.14 $\pm$    0.05 &    0.27 $\pm$    0.04\\[\cmsTabSkip]
10 & $>$1500                     & $>$1500    &    3.54 $\pm$    0.84 &    3.75 $\pm$    0.72 &    1 &    0.08 $\pm$    0.04 &    0.03 $\pm$    0.01\\
\hline
\end{tabular}
}
\label{tab:eventyields}
\end{table*}

In addition to the background and data yields, two benchmark signal points are shown. The first one is an SM1 signal with $\mSmu = 0.4\TeV$ and a neutralino mass of $\mNeutr = 0.2\TeV$. It is normalized to a cross section of 13.8\unit{fb}, which corresponds to a coupling of $\lpmuo = 0.0016$ in the modified cMSSM for this process and the chosen masses. The second signal benchmark, from SM2, is normalized to a cross section of 1.20\unit{fb}, corresponding to $\lpmuo = 0.0088$. The corresponding slepton mass is 1.4\TeV, the neutralino mass is 0.5\TeV, and $x=0.5$.
The combined acceptance times efficiency is 11\% and 31\% for the first and second benchmark signal points, respectively.

\begin{figure*}
\centering
\includegraphics[width=\cmsFigWidth]{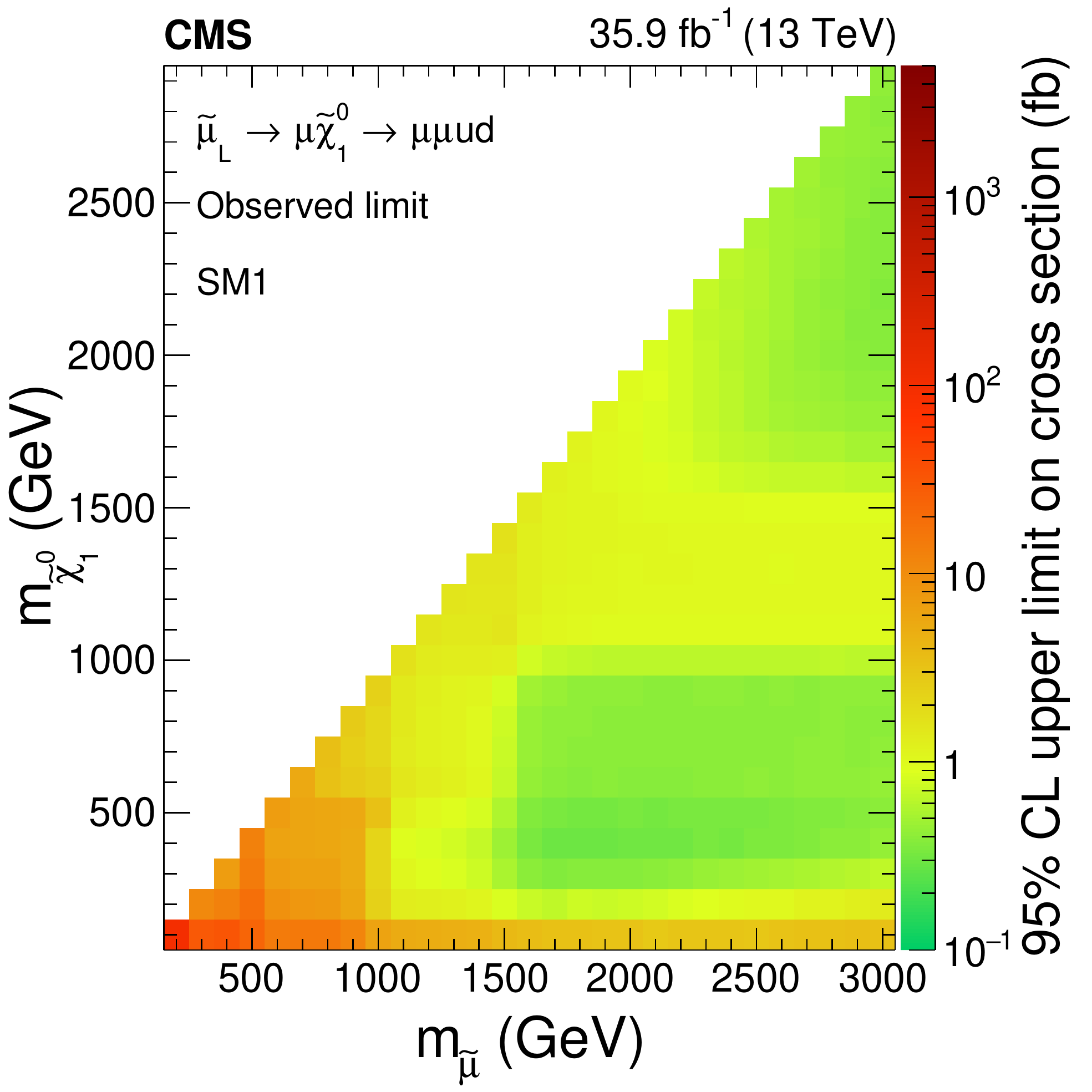}
\includegraphics[width=\cmsFigWidth]{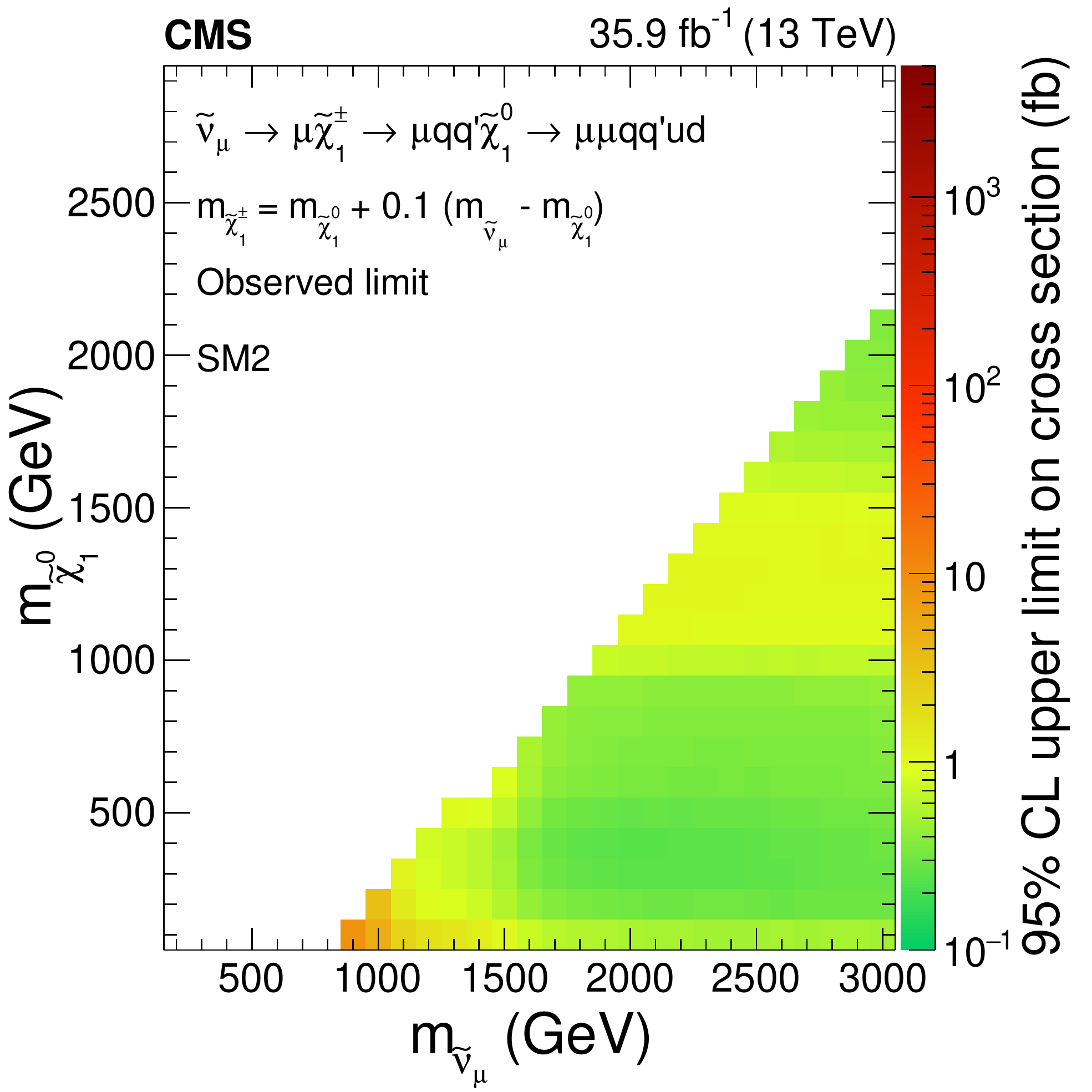}\\
\includegraphics[width=\cmsFigWidth]{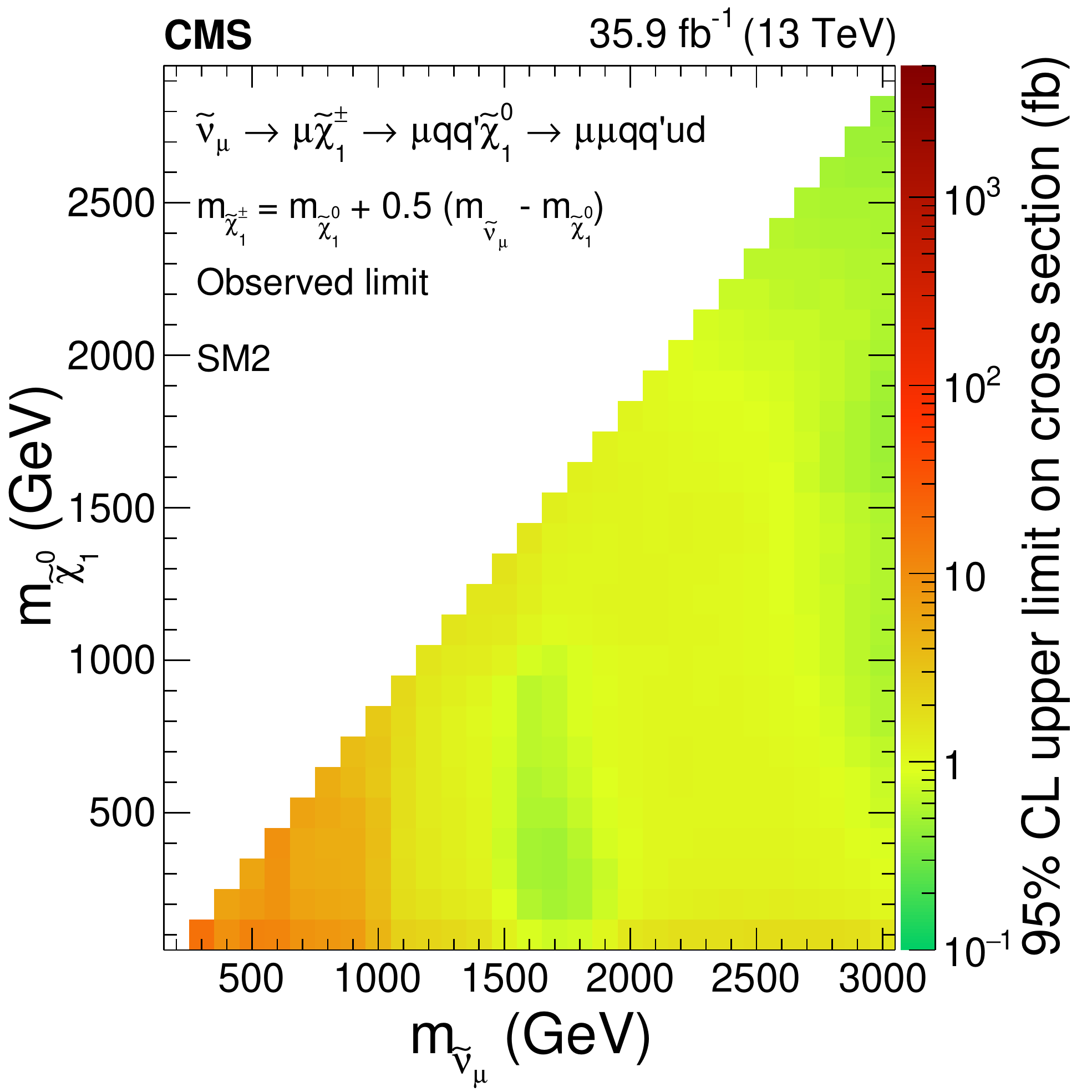}
\includegraphics[width=\cmsFigWidth]{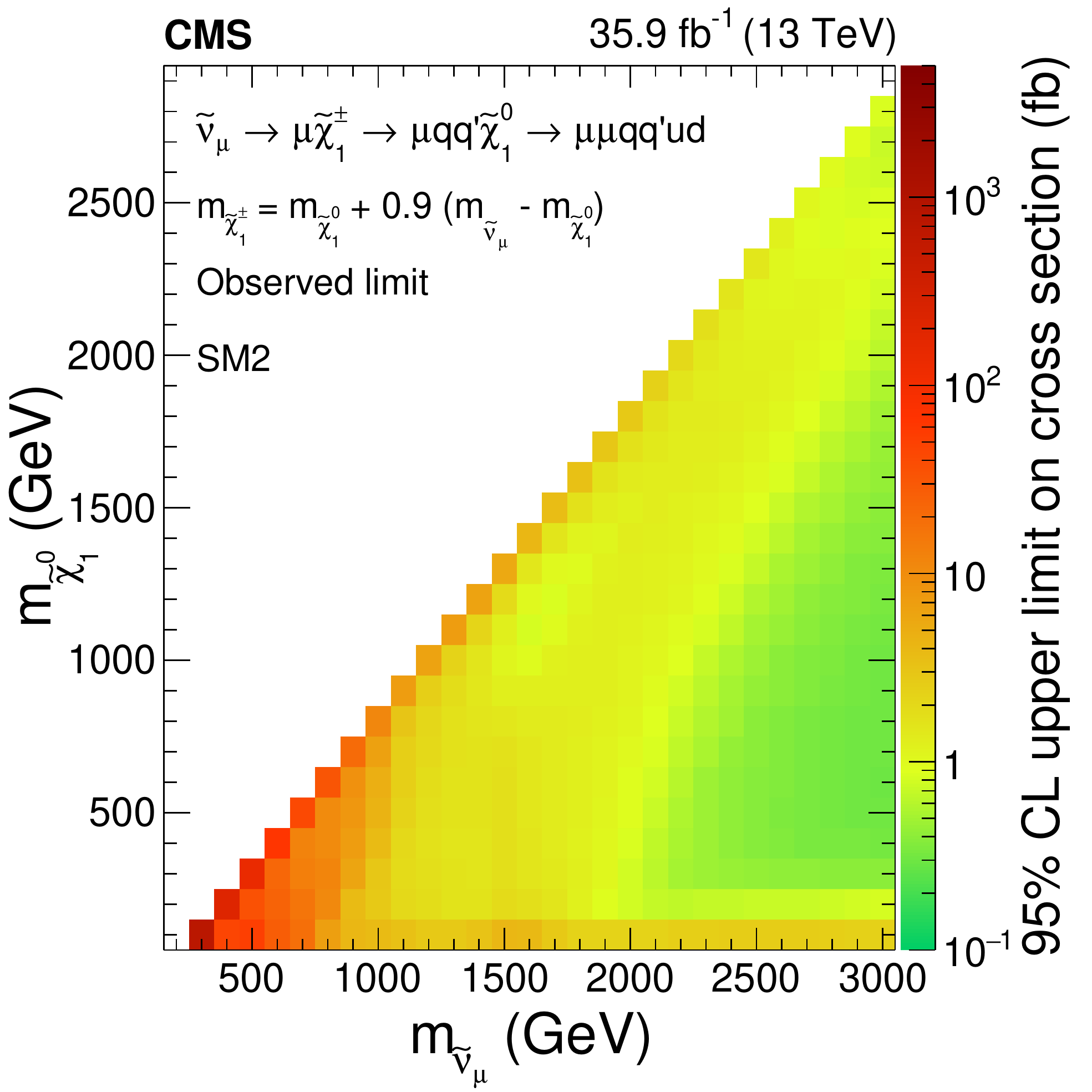}
\caption{Observed upper limits on cross sections at 95\% \CL. The upper left plot shows the limit in the \mNeutr{} and \mSmu{} mass plane for SM1, while the other three plots show the SM2 limits as a function of \mNeutr{} and \mSnuMu{} for the three different scenarios with $x = 0.1$ (upper right), $x=0.5$ (lower left) and $x=0.9$ (lower right). The limit for a specific mass combination is depicted according to the color scale on the right-hand side of the figures.}
\label{fig:xslimit_sm}
\end{figure*}

The results are interpreted in terms of the simplified models introduced in Section \ref{sec:introduction}.
Upper limits on cross sections are set at 95\% confidence level (\CL) using the \CLs criterion~\cite{Junk:1999kv,Read:2002hq,ATLAS:1379837} in the asymptotic approximation~\cite{Cowan:2010js} with the frequentist profile likelihood ratio presented in Ref.~\cite{ATLAS:1379837}.
The uncertainties listed in Section \ref{sec:unc} are included as nuisance parameters assuming log-normal distributions and are profiled in the limit setting.
The observed cross section upper limits are shown in Fig.~\ref{fig:xslimit_sm} as a function of the sparticle masses of each signal point.

\begin{figure}[!htb]
\centering
\includegraphics[width=\cmsFigWidth]{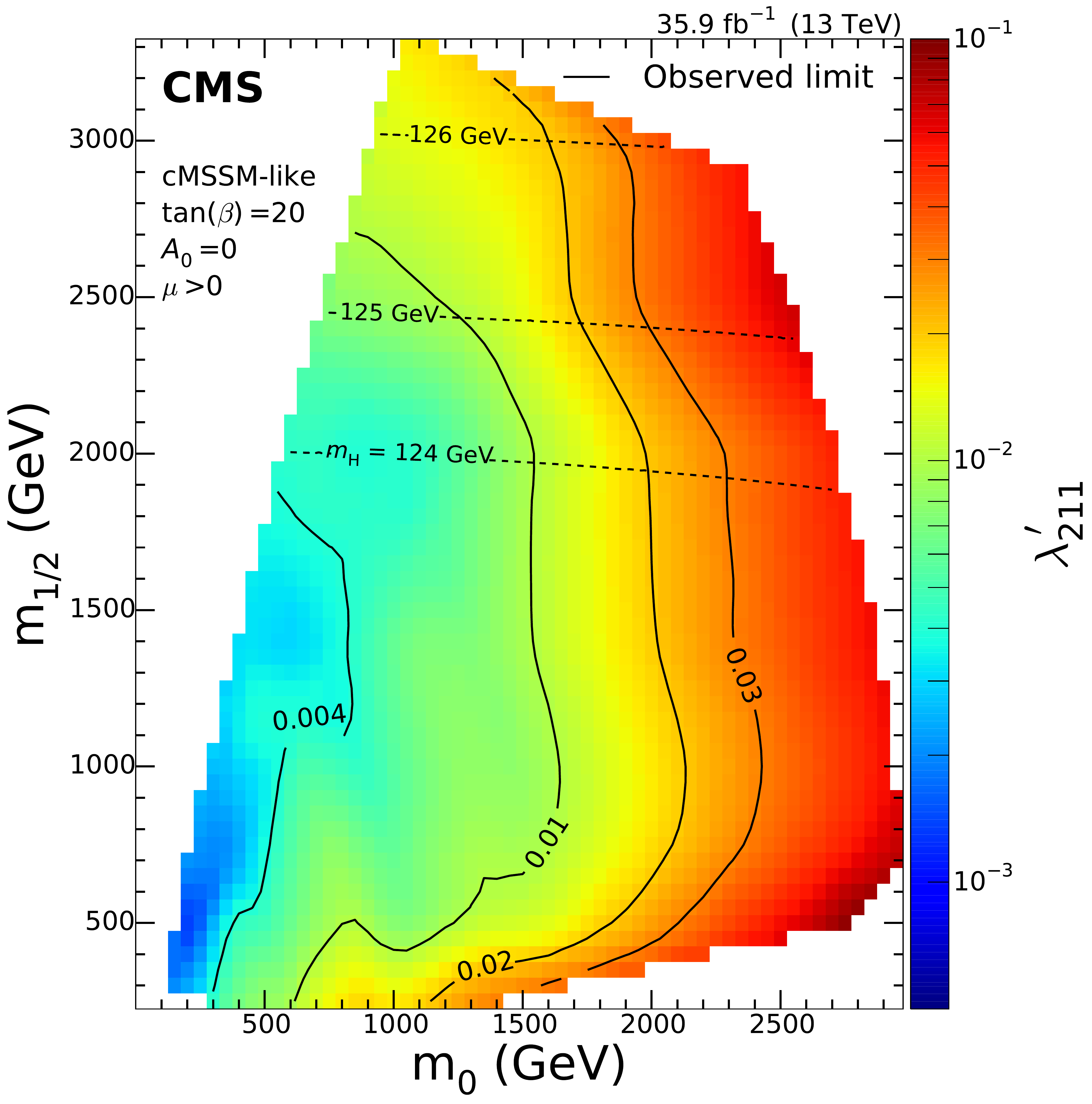}
\caption{Upper limits at 95\% \CL on the coupling \lpmuo as a function of \mzero and \monehalf for a modified cMSSM with \lpmuo as additional RPV coupling. The color scale at the right side of the figure indicates the coupling limit value for specific parameter combinations. These limits are derived from the upper cross section limits of SM1. For four values of \lpmuo (0.004, 0.01, 0.02, 0.03), the coupling limits are shown as black contour lines. The dashed lines show the parameters in the model that correspond to the mass of the lightest Higgs boson for three chosen values (124, 125, 126\GeV).}
\label{fig:lplimit}
\end{figure}

The upper bounds on cross sections are translated into coupling limits of the full cMSSM-like model with an additional RPV coupling \lpmuo as explained in Section~\ref{sec:introduction}. For this benchmark model, the cMSSM parameters are set to $\tan\beta = 20$, $\mu > 0$, and $A_0 = 0$. Here, $\tan\beta$ is the ratio of the vacuum expectation values of the neutral components of the two Higgs doublets, $\mu$ the SUSY Higgsino mass parameter, and $A_0$ the universal trilinear coupling.  The coupling limits are derived for each mass combination of \PSGmL and \PSGczDo in SM1 where the mass combination corresponds to a valid cMSSM point. The full model cross section times the branching fraction for the decay according to SM1 is equal to the observed SM1 cross section limit at a specific value of \lpmuo. This value corresponds to the expected upper bound on the coupling. Full model cross sections have been calculated with \MGvATNLO for a coupling value of $\lpmuo = 0.01$. All \lpmuo coupling values are given at the unification scale. Cross sections for different values of the coupling are extrapolated assuming a scaling of the cross section with $\lambda_{211}^{\prime 2}$. Signal points where this assumption is not valid are discarded, \eg, for values where the branching fraction of the \PSGmL or \sNuMu into quarks becomes relevant. The resulting \lpmuo limits based on SM1 are shown in Fig.~\ref{fig:lplimit} as a function of \mzero and \monehalf, denoting the universal scalar and gaugino masses in the cMSSM, respectively. For the cMSSM-like model, no constraint on the Higgs boson mass was imposed. For three chosen values, the parameters corresponding to the mass of the lightest Higgs boson in the model calculated with a top quark mass of 172.5\GeV are shown as dashed lines. Using a similar method, coupling limits are derived for the SM2 points where the three involved masses correspond to a valid cMSSM point. These results are listed in Table~\ref{tab:limits}. For the scan with $x=0.9$, no point matches the criteria above.

\begin{table*}[!htb]
\centering
\topcaption{Observed upper limits on cross sections at 95\% \CL for selected SM2 points. The corresponding limits on \lpmuo for the modified cMSSM with \lpmuo as additional coupling are shown as well.}
\begin{tabular}{ccccccc}
\hline
\mzero ({\GeVns}) & \monehalf ({\GeVns}) & \mSnuMu ({\GeVns}) & \mNeutr ({\GeVns}) & $x$ & Cross section limit (fb) & \lpmuo limit \\
\hline
 890 &  250 &  900 &  100 & 0.1 & 8.7 & 0.0085 \\
 990 &  250 & 1000 &  100 & 0.1 & 5.0 & 0.0081 \\
1880 &  480 & 1900 &  200 & 0.1 & 0.32 & 0.0093 \\
1980 &  480 & 2000 &  200 & 0.1 & 0.31 & 0.011 \\
2670 &  700 & 2700 &  300 & 0.1 & 0.27 & 0.026 \\
2770 &  700 & 2800 &  300 & 0.1 & 0.28 & 0.031 \\
1180 & 1160 & 1400 &  500 & 0.5 & 1.08 & 0.0084 \\
1860 & 1820 & 2200 &  800 & 0.5 & 1.05 & 0.028 \\
2280 & 2250 & 2700 & 1000 & 0.5 & 0.84 & 0.048 \\
2550 & 2470 & 3000 & 1100 & 0.5 & 0.57 & 0.058 \\
\hline
\end{tabular}
\label{tab:limits}
\end{table*}

\section{Summary}
A search for resonant production of second-generation sleptons (\PSGmL, \sNuMu) using 35.9\fbinv of proton-proton collisions recorded in 2016 with the CMS detector has been presented. The search targets resonant slepton production via the $R$-parity violating coupling \lpmuo to quarks in final states with two same-sign muons and at least two jets. No significant excess over the background expectation is observed.
Upper limits on cross sections are set in the context of two simplified models covering the dominant production mechanisms in a modified constrained minimal supersymmetric model (cMSSM) with \lpmuo as an additional coupling. These limits, ranging from 0.24 to 730\unit{fb}, are translated into limits on the coupling \lpmuo in the modified cMSSM, and represent the most stringent limits on this particular model of $R$-parity violating supersymmetry.

\begin{acknowledgments}
We congratulate our colleagues in the CERN accelerator departments for the excellent performance of the LHC and thank the technical and administrative staffs at CERN and at other CMS institutes for their contributions to the success of the CMS effort. In addition, we gratefully acknowledge the computing centres and personnel of the Worldwide LHC Computing Grid for delivering so effectively the computing infrastructure essential to our analyses. Finally, we acknowledge the enduring support for the construction and operation of the LHC and the CMS detector provided by the following funding agencies: BMBWF and FWF (Austria); FNRS and FWO (Belgium); CNPq, CAPES, FAPERJ, FAPERGS, and FAPESP (Brazil); MES (Bulgaria); CERN; CAS, MoST, and NSFC (China); COLCIENCIAS (Colombia); MSES and CSF (Croatia); RPF (Cyprus); SENESCYT (Ecuador); MoER, ERC IUT, and ERDF (Estonia); Academy of Finland, MEC, and HIP (Finland); CEA and CNRS/IN2P3 (France); BMBF, DFG, and HGF (Germany); GSRT (Greece); NKFIA (Hungary); DAE and DST (India); IPM (Iran); SFI (Ireland); INFN (Italy); MSIP and NRF (Republic of Korea); MES (Latvia); LAS (Lithuania); MOE and UM (Malaysia); BUAP, CINVESTAV, CONACYT, LNS, SEP, and UASLP-FAI (Mexico); MOS (Montenegro); MBIE (New Zealand); PAEC (Pakistan); MSHE and NSC (Poland); FCT (Portugal); JINR (Dubna); MON, RosAtom, RAS, RFBR, and NRC KI (Russia); MESTD (Serbia); SEIDI, CPAN, PCTI, and FEDER (Spain); MOSTR (Sri Lanka); Swiss Funding Agencies (Switzerland); MST (Taipei); ThEPCenter, IPST, STAR, and NSTDA (Thailand); TUBITAK and TAEK (Turkey); NASU and SFFR (Ukraine); STFC (United Kingdom); DOE and NSF (USA).

\hyphenation{Rachada-pisek} Individuals have received support from the Marie-Curie programme and the European Research Council and Horizon 2020 Grant, contract No. 675440 (European Union); the Leventis Foundation; the A. P. Sloan Foundation; the Alexander von Humboldt Foundation; the Belgian Federal Science Policy Office; the Fonds pour la Formation \`a la Recherche dans l'Industrie et dans l'Agriculture (FRIA-Belgium); the Agentschap voor Innovatie door Wetenschap en Technologie (IWT-Belgium); the F.R.S.-FNRS and FWO (Belgium) under the ``Excellence of Science - EOS" - be.h project n. 30820817; the Ministry of Education, Youth and Sports (MEYS) of the Czech Republic; the Lend\"ulet (``Momentum") Programme and the J\'anos Bolyai Research Scholarship of the Hungarian Academy of Sciences, the New National Excellence Program \'UNKP, the NKFIA research grants 123842, 123959, 124845, 124850 and 125105 (Hungary); the Council of Science and Industrial Research, India; the HOMING PLUS programme of the Foundation for Polish Science, cofinanced from European Union, Regional Development Fund, the Mobility Plus programme of the Ministry of Science and Higher Education, the National Science Center (Poland), contracts Harmonia 2014/14/M/ST2/00428, Opus 2014/13/B/ST2/02543, 2014/15/B/ST2/03998, and 2015/19/B/ST2/02861, Sonata-bis 2012/07/E/ST2/01406; the National Priorities Research Program by Qatar National Research Fund; the Programa Estatal de Fomento de la Investigaci{\'o}n Cient{\'i}fica y T{\'e}cnica de Excelencia Mar\'{\i}a de Maeztu, grant MDM-2015-0509 and the Programa Severo Ochoa del Principado de Asturias; the Thalis and Aristeia programmes cofinanced by EU-ESF and the Greek NSRF; the Rachadapisek Sompot Fund for Postdoctoral Fellowship, Chulalongkorn University and the Chulalongkorn Academic into Its 2nd Century Project Advancement Project (Thailand); the Welch Foundation, contract C-1845; and the Weston Havens Foundation (USA).
\end{acknowledgments}

\bibliography{auto_generated}
\cleardoublepage \appendix\section{The CMS Collaboration \label{app:collab}}\begin{sloppypar}\hyphenpenalty=5000\widowpenalty=500\clubpenalty=5000\vskip\cmsinstskip
\textbf{Yerevan Physics Institute, Yerevan, Armenia}\\*[0pt]
A.M.~Sirunyan, A.~Tumasyan
\vskip\cmsinstskip
\textbf{Institut f\"{u}r Hochenergiephysik, Wien, Austria}\\*[0pt]
W.~Adam, F.~Ambrogi, E.~Asilar, T.~Bergauer, J.~Brandstetter, M.~Dragicevic, J.~Er\"{o}, A.~Escalante~Del~Valle, M.~Flechl, R.~Fr\"{u}hwirth\cmsAuthorMark{1}, V.M.~Ghete, J.~Hrubec, M.~Jeitler\cmsAuthorMark{1}, N.~Krammer, I.~Kr\"{a}tschmer, D.~Liko, T.~Madlener, I.~Mikulec, N.~Rad, H.~Rohringer, J.~Schieck\cmsAuthorMark{1}, R.~Sch\"{o}fbeck, M.~Spanring, D.~Spitzbart, A.~Taurok, W.~Waltenberger, J.~Wittmann, C.-E.~Wulz\cmsAuthorMark{1}, M.~Zarucki
% ...
\vskip\cmsinstskip
% ...
1:  Also at Vienna University of Technology, Vienna, Austria\\
% ...
\end{sloppypar}
\end{document}